\documentclass[journal=jacsat,manuscript=article]{achemso}
\usepackage{braket}
\usepackage[version=3]{mhchem} 
\usepackage[usenames,dvipsnames]{color}

\author{Emily G. Ward}
\affiliation{Department of Chemistry, Indiana University, Bloomington, IN 47405-7102, USA}
\author{Alexandru B Georgescu}
\affiliation{Department of Chemistry, Indiana University, Bloomington, IN 47405-7102, USA}
\email{georgesc@iu.edu}

\title[An \textsf{achemso} demo]
  {Tight-Binding Models for Lone Pair, Heteroanionic Solids, and Application to Layered Oxyhalides}

\abbreviations{IR,NMR,UV}

\begin{document}




\begin{abstract}
We provide a methodology to understand materials with complex bonding patterns, and apply it to the example of heteroanionic and lone pair materials. We build a tight-binding model based on Wannier functions fitted on density functional theory results, followed by enforcing symmetry on the atomic orbital basis set, and finally connecting and disconnecting sets of orbitals from the tight-binding model to understand their individual contribution to the resulting materials properties. We apply this methodology to complex materials, namely \ce{BiOCl} and \ce{Bi$_2$YO$_4$Cl} - part of a broader class of materials investigated for their applications in photocatalysis and photoluminescence. Our methodology can be generalized and applied to a wide variety of other materials, including halide perovskites, and multiferroic materials. This methodology allows us to isolate the origin of key electronic features in these materials, including the role of the Bi lone pair-anion bonding interaction - key to photoluminescence in many materials. Finally, we investigate the role of the crystal structure, Chlorine and Oxygen orbital energy levels and bonding in determining the photostability of bismuth oxyhalides. Our methodology allows us to understand the functionality of complex materials in an intuitive and qualitative manner.

\end{abstract}

\section{Introduction}

Key to building predictive theories for the rational design of novel materials is understanding their complex bonding patterns. Tight-binding models, which are effective single particle band models containing local crystal field energies and 'hoppings' between different atoms, can provide key insight. Historically, Linear Combinations of Atomic Orbitals (LCAO) methods have been used to build effective tight-binding models in solids, however, with the advent of ab-initio plane-wave density functional theory, it is more common to use standard tools for electronic structure analysis, including projected density of states, and orbitally-projected bands. As orbital projectors provided with DFT codes and/or pseudopotentials generally do not provide a complete, orthogonal and normalized set of states to represent a particular solid, an alternative is provided by Wannier function-based models. Wannier functions form a complete, orthogonal, and normalized set of states that describe electronic states within a specified energy range\cite{wannier1,wannier2}. However, they come with their own limitations, as multiple sets of Wannier functions can reproduce the band structure within a certain energy range. The procedure to obtain them usually requires minimization of the Wannier function spread, while reproducing a subset of the DFT bands exactly, and may depend on the initial starting seed - a procedure referred to as Maximally Localized Wannier Functions (MLWF)\cite{MLWF,MLWF2}. The resulting model and local orbital basis can also be used for additional quantum many-electron calculations that require a specific local basis, most commonly for correlated electron materials\citep{DMFT1,DMFT2,DMFT3,DMFT4}. Depending on the initial parameters, energy windows, symmetry conditions, and minimization of the Wannier function spread, Wannier models may differ, and make it difficult to interpret physical meaning. 

Lone pairs control many properties of many solid materials, including electric polarization, multiferroic behavior, or photocatalytic and photoluminescence mechanisms\cite{Zhong2022,RamSeshradiLP,Walsh2,SpaldinRamLP,LP2}. Lone pairs are an electron pair in a filled orbital. Typically, in solids, they are hybrids of s and p orbitals that arise as local inversion symmetry is broken and eigenstates are no longer odd or even along a certain direction. They are often considered to be chemically inactive - and only stereochemically active - with their role often underestimated\cite{RamSeshradiLP,Fabini_Seshadri_Kanatzidis_2020,Morgan2020}. Modeling their exact contribution to physical and chemical properties can be difficult, particularly to obtain a quantitative theory. While Valence Shell Electron Pair Repulsion (VSEPR) theory is commonly used to qualitatively understand how lone pairs affect the coordination geometry of a molecule, this can provide limited intuition in solids - let alone a quantitative interpretation of observed phenomena. A computational method that allows visualization and identification of lone pairs, and bonding in solids more generally, is the Electron Localization Function (ELF)\cite{ELFRef,ELF2,ELF3}, which can show where the electrons are localized in a material. However, while these methods are useful, it is key to build new quantitative approaches to understand the complex bonding interactions in these classes of materials\cite{LPReview}.

We propose a methodology to understand complex bonding patterns in solids, and apply it to the study of lone pairs and multiple anions in solid materials, using tight-binding models built from Wannier functions. After performing DFT calculations, followed by a Wannierization procedure, we use a local diagonalization procedure, building on what was originally described by Georgescu and coworkers\cite{PrevLPABG} for 2D van der Waals materials. This allows us to separate the hybridized s-p orbitals, namely the s-like lone pair from the p orbitals on the Bi 3+ ion. After obtaining a tight-binding model, we can further quantify the effect of each orbital on the electronic structure of the material by disconnecting it from the other orbitals by simply setting the hopping terms (which in Hamiltonian notation correspond to $\braket{ \Psi_1 | \hat{H} |\Psi_2 }$ for hopping from an orbital 1 to orbital 2) to 0. This then allows us to directly visualize the key orbitals, including the Bi lone pair orbitals. We demonstrate the utility of this method by analyzing the lone pairs of bismuth oxyhalide materials, which are commonly studied for photocatalytic and photoluminescent applications. We provide the resulting scripts on Github\cite{Github}. 

Specifically, we studied effect of the lone pairs in \ce{BiOCl} and \ce{Bi2YO4Cl}, as they are part of a wider materials' family of materials where lone pairs play a key role in photocatalytic and photoluminescent properties\cite{Bi2MO4ClLP,Kaustav,Synthesis,optoelectronic,Lone,Optical,Oxyhalides2,Oxyhalides3,Oxyhalides4}. Both systems are layered, multi-anion materials with lone pairs located on their \ce{Bi^{3+}} ions with similar symmetries (Table \ref{tab:symmetrydata})\cite{BiOClMaterialsProject,Bi2YO4ClMaterialsProject,MaterialsProject}. Previous works have interpreted the electronic structure of  bismuth oxyhalides using Madelung potentials to characterize on-site atomic energies and using Revised Lone-Pair Theory for qualitative insight into orbital energies\cite{madelung1,madelung2,RevisedLPTheory,Bi2MO4ClLP}. Our study can provide direct quantitative insight into the interactions between specific orbitals, and the resulting phenomenology. 

\section{Application to \ce{BiOCl} and \ce{Bi2YO4Cl}}
\subsection{Lone Pair Visualization and Analysis}
\begin{figure}
    \centering
    \includegraphics[width=1\linewidth]{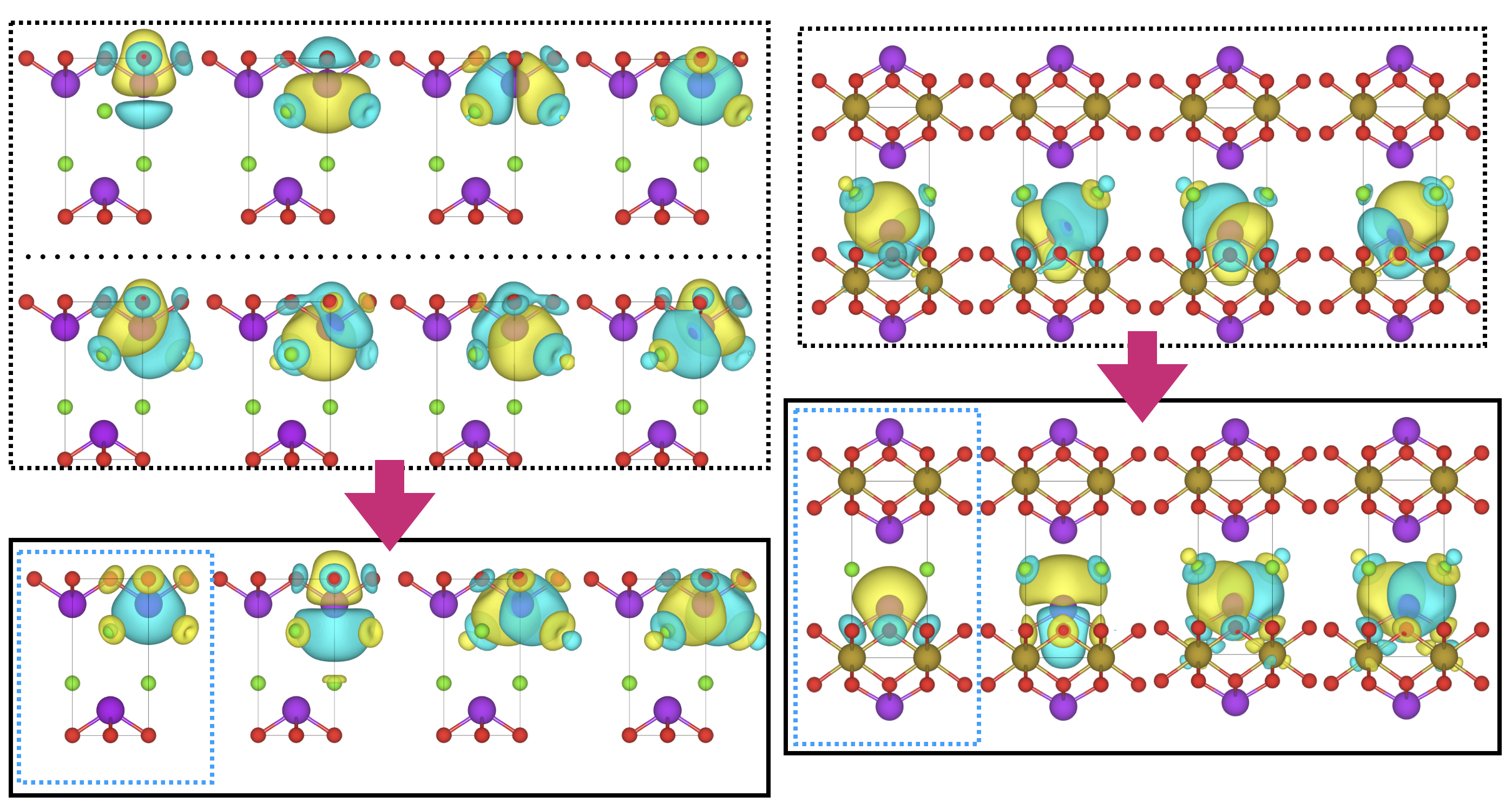}
    \caption{Left: Initial MLWF obtained from s and p seeds, and from sp$^3$ seeds, and the resulting Wannier functions following our Hamiltonian rotation procedure - which do not depend on the initial guess. Right: \ce{Bi2YO4Cl} MLWF before rotation (sp$^3$ and after (below) Hamiltonian rotation.}
    \label{fig:wanniers}
\end{figure}

\begin{figure}
    \centering
    \includegraphics[width=1\linewidth]{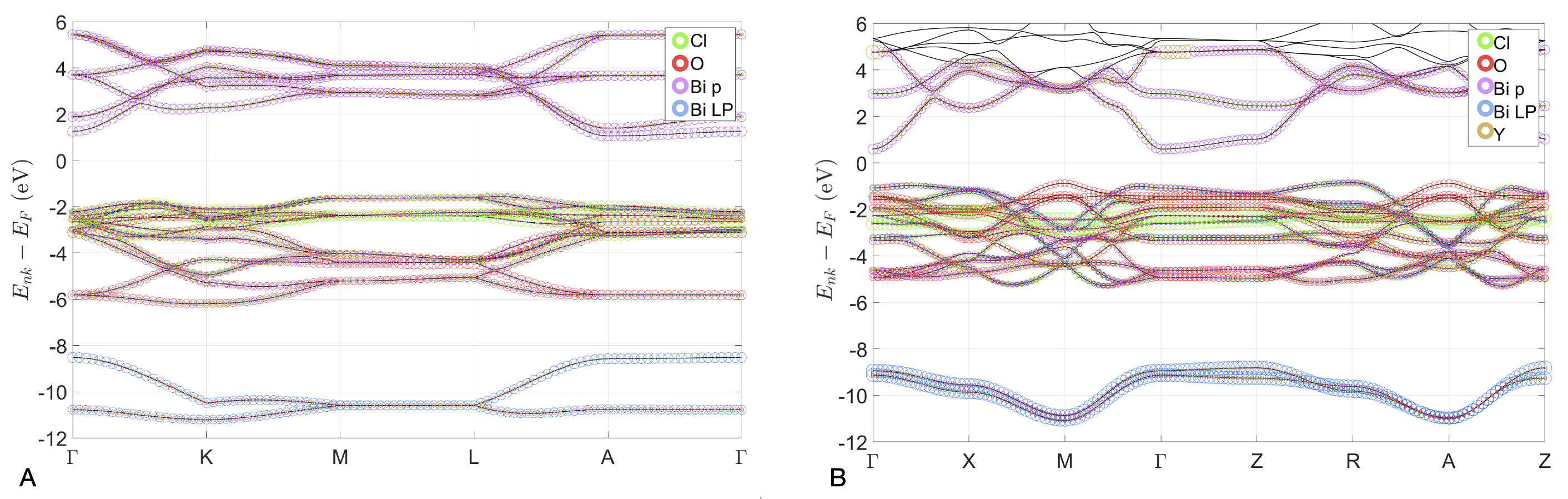}
    \caption{Projected bands based on the Wannier Hamiltonian for \ce{BiOCl} (A) and \ce{Bi2YO4Cl} (B), which fully reproduce the DFT bands (black lines). Despite being around 10eV below the Fermi level, the Bi lone pair state has a significant contribution near the top of the conduction band maximum.}
    \label{fig:Bands}
\end{figure}
 We illustrate the usefulness of our methodology by applying it to two example complex materials. The first step in the procedure is to fully relax the crystal structures of the two materials (\ce{BiOCl} and \ce{Bi2YO4Cl}) within density functional theory. We then use the Wannier90 code to obtain maximally localized Wannier functions that fully reproduce the DFT bands from around -12eV below, to around 7 eV above the Fermi energy, and their corresponding tight-binding Hamiltonian containing local crystal field energies and 'hoppings' between the orbitals. This procedure is made more efficient and more likely to give physically meaningful results by the use of an initial 'seed' corresponding to the nature of the orbitals we expect in the material. We use p orbitals as an initial seed for the Cl and O atoms, the two $e_g$ d-orbitals for the Y atom, and either the sp$^3$ or s and p orbitals for the Bi atom. We verify that the tight binding model we obtain reproduces the DFT band structure (Figure \ref{fig:Bands}). After obtaining this initial set of Wannier functions, we can rotate the Hamiltonian via a linear, unitary transformation, such that the on-site Hamiltonian on the Bi site becomes diagonal, leading to 0 inter-orbital hoppings on each Bi atom; our results do not depend on whether the initial seed consisted of the sp$^3$ or s and p orbitals for the Bi. This procedure is not necessary on the other atoms (though it is possible) as the basis obtained from the Wannier procedure is nearly diagonal to begin with.

We then obtain an orbital set that naturally conforms to the local symmetry of the system (Figure \ref{fig:wanniers}). A similar methodology has been applied in the case of 2D magnetic halides (MX$_2$, MX$_3$, with M a transition metal and X={Cl,Br,I}), where it was used to easily explain their insulating nature using a local trigonal basis for the d-orbitals. Here, this rotation allows us to obtain a Hamiltonian clearly separating the Bi lone pair, and the Bi p-orbitals. The initial choice of seed for our Wannierization procedure for the Bi ion can include either s and p orbitals, or four sp$^3$ orbitals, leading to two different sets of Wannier functions. However, after the initial Wannier functions have been generated, the resulting lone pair orbital and p-like orbitals are independent of the initial seed, and so is the resulting Hamiltonian and orbitals - as plotted in Figure \ref{fig:wanniers}. The diagonalization procedure creates a transformation matrix for each atomic orbital that will transform the original orbital basis set into a diagonal orbital basis set in increasing crystal field energy order. As depicted in Figure \ref{fig:wanniers}, we can visualize  the lone pair orbital location, shape and phase on the bismuth atom in each material.

\begin{figure}
    \centering
    \includegraphics[width=1\linewidth]{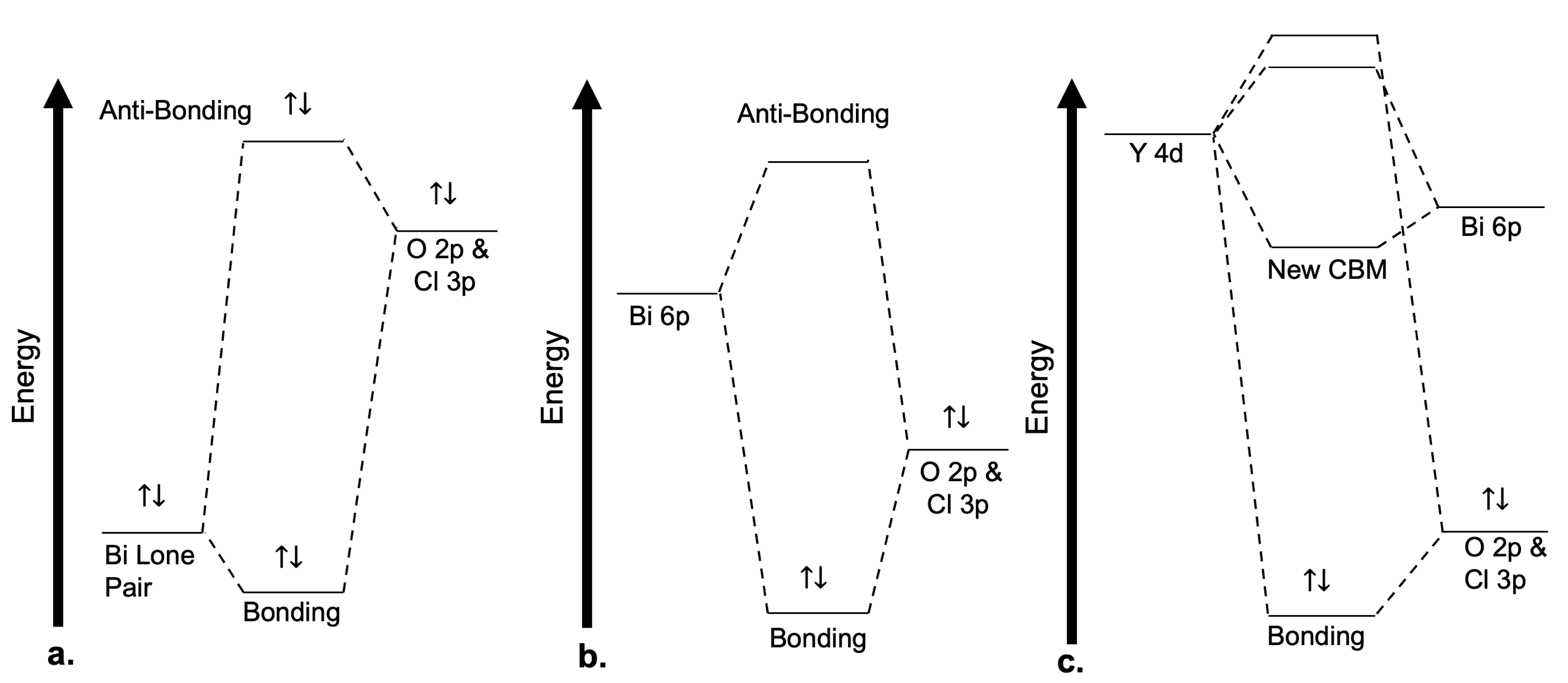}
    \caption{Summary of bonding interactions in the two materials. The initial relative energy levels of the orbitals are partially set by Madelung potentials. While this sketch describes the formation of bonding/antibonding pairs, it does not appropriately capture effects on the bandwidths corresponding to the different orbitals, and the sequential effect of each bonding interaction. Of these interactions, Y d orbital bonding has the lowest effect on the band gap.}
    \label{fig:Bonding}
\end{figure}

Our method provides a standardized method for creating a tight binding model including a lone pair orbital within a crystal structure, as well as a method to quantitatively analyze its role and that of other orbitals in a systematic way. We summarize the main interactions in Figure \ref{fig:Bonding}; table \ref{tab:BandGaps} summarizes the effect of connecting/disconnecting different orbitals on the band gap. 

\begin{figure}
    \centering
    \includegraphics[width=1\linewidth]{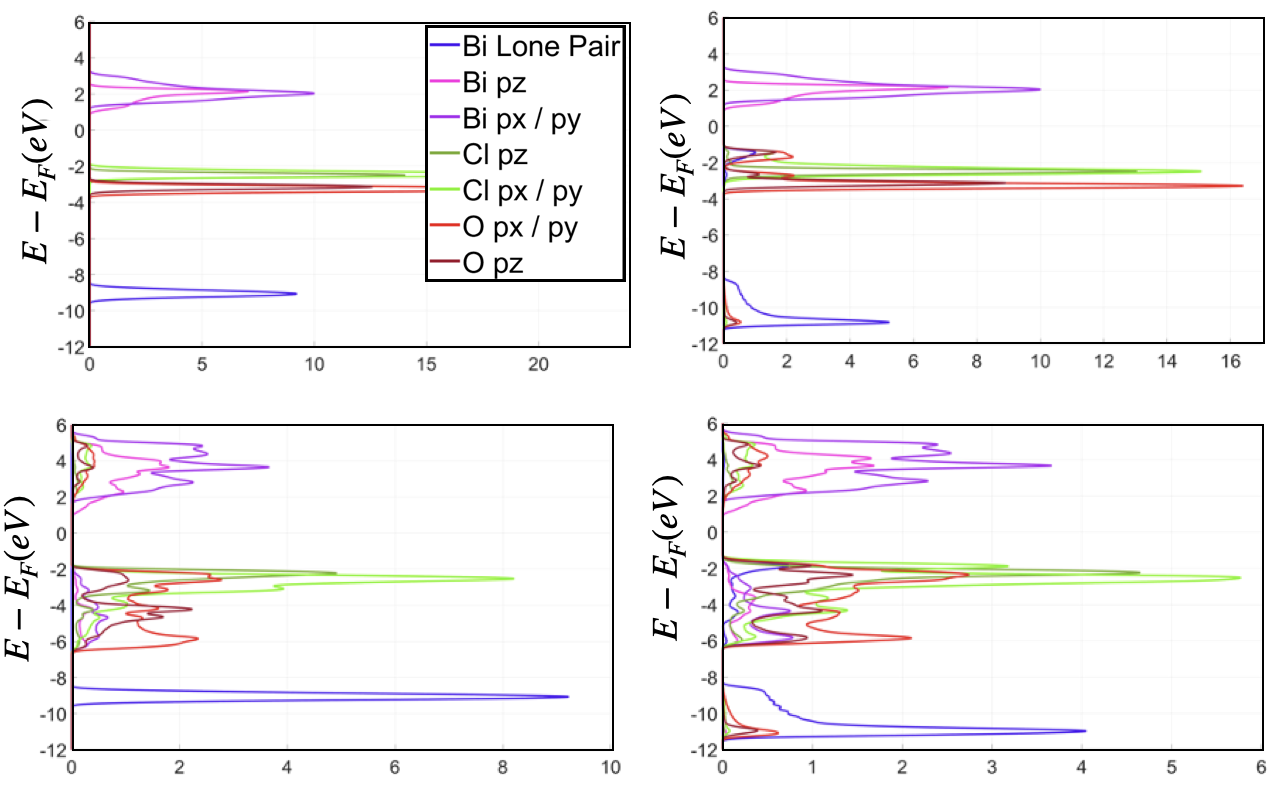}
    \caption{Projected Density of States (PDOS) as obtained from the Wannier Hamiltonians for \ce{BiOCl}. A: Hoppings set to 0. Diagonal elements of the hopping from one unit cell to another for R nonzero allowed, leaving a small bandwidth. B: Only lone pair hoppings allowed. C: All but lone pair hoppings are allowed. D: PDOS for the full Hamiltonian.  }
    \label{fig:PDOS}
\end{figure}

\begin{figure}
    \centering
    \includegraphics[width=1\linewidth]{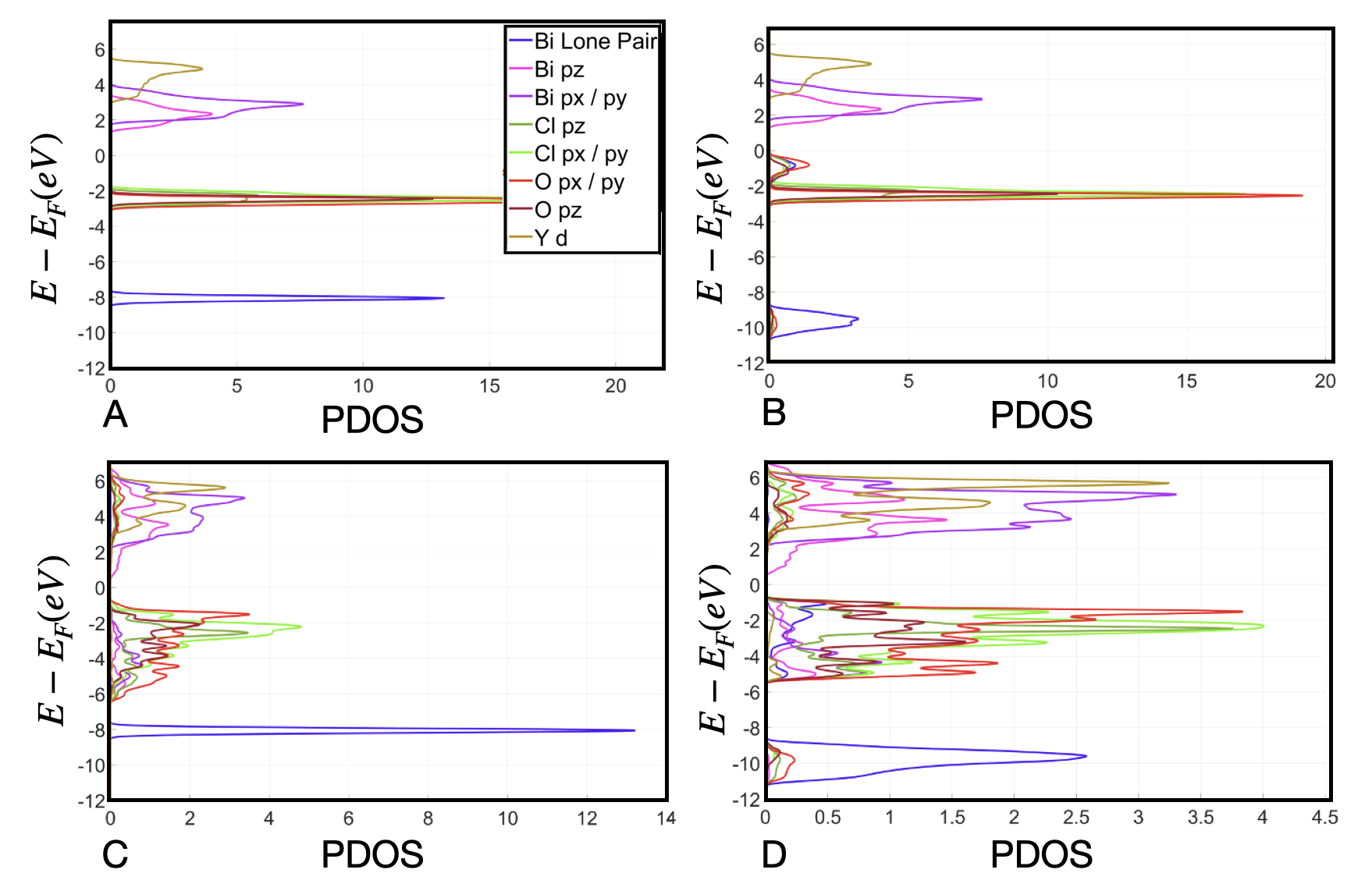}
    \caption{Projected Density of States (PDOS) as obtained from the Wannier Hamiltonians for \ce{Bi2YO4Cl}. A: Hoppings set to 0. Diagonal elements of the hopping from one unit cell to another for R nonzero allowed, leaving a small bandwidth. B: Only lone pair hoppings allowed. C: All but lone pair hoppings are allowed. D: PDOS for the full Hamiltonian.  }
    \label{fig:PDOS2}
\end{figure}

As the resulting Hamiltonian has many entries (for example, the $\vec{R}=[0,0,0]$ component of the Hamiltonian - hopping within the same unit cell  - for BiOCl corresponds to a 20x20 matrix with 400 entries, with many non-zero), it can be difficult to obtain an intuitive understanding of the bonding in this material. What we propose in this paper is to connect or disconnect the orbitals from the tight-binding Hamiltonian by setting specific hoppings to 0. This can be done in two ways: one is to disconnect all orbitals and selectively connect them back, and the other is to start from the full tight-binding Hamiltonian and disconnect select orbitals. 

Each aspect of this procedure allows us to gain direct insight into different aspects of the electronic structure of these materials, despite their relatively high complexity.

\begin{figure}
    \centering
    \includegraphics[width=1\linewidth]{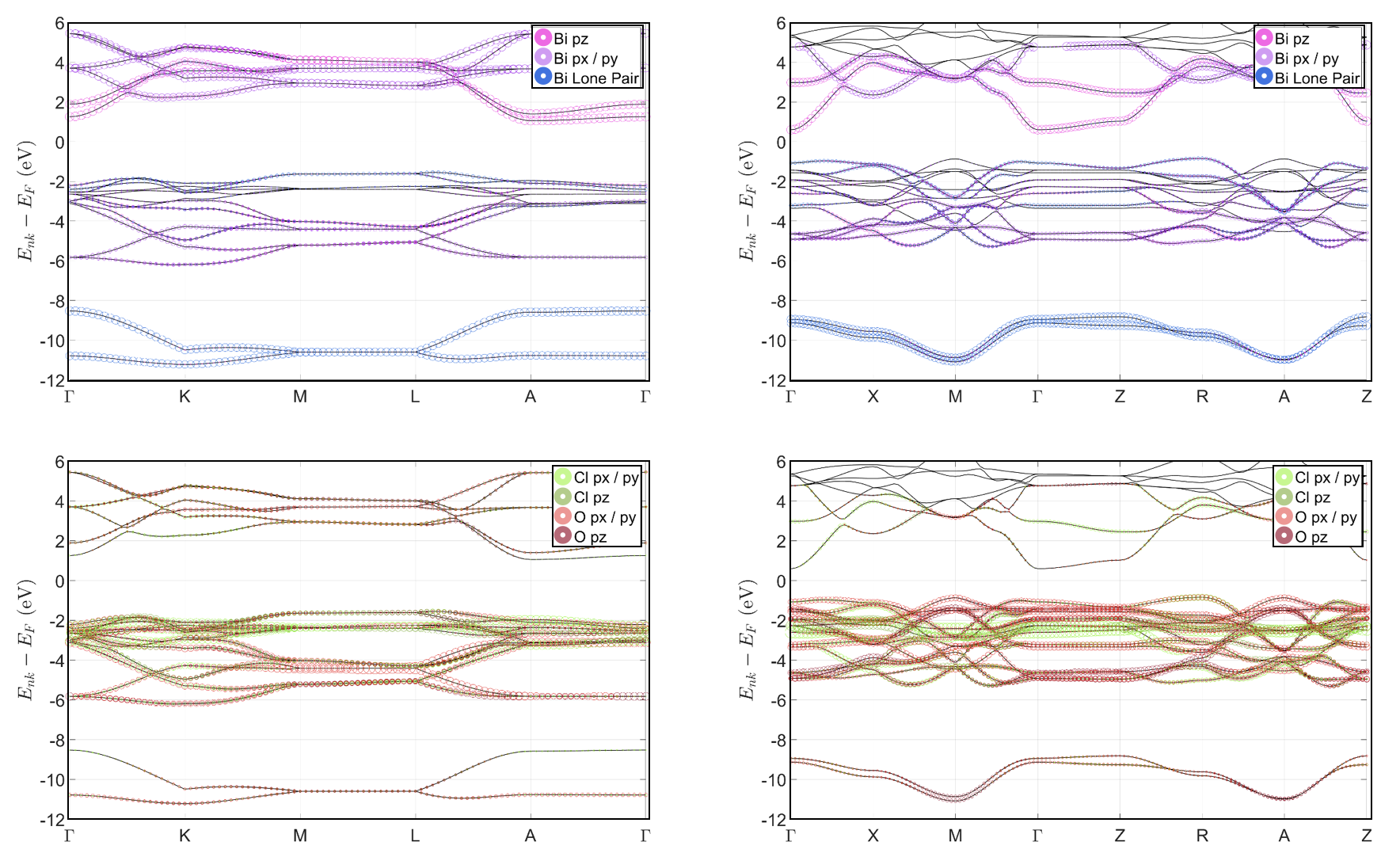}
    \caption{Orbitally projected bands from the Wannier Hamiltonian on the DFT bands. Left: BiOCl, Right: \ce{Bi2YO4Cl}. The Bi bands (top) display lone pair character at the conduction band maximum, key to photoluminescence in this class of materials, and in other classes of materials containing lone pairs. Bottom: the relative arrangement of the Cl and O p orbitals is key to the photostability of  \ce{Bi2YO4Cl} relative to BiOCl and its possible applications in photocatalysis, with  \ce{Bi2YO4Cl} showing mostly O p character at the VBM, while BiOCl mostly Cl.}
    \label{fig:BandsDetailed}
\end{figure}

For example, layered oxyhalides have been studied for their photoluminescence and photocatalytic properties, both of which are closely linked to their unique electronic structure, and the role of the lone pairs in the electronic structure. Our Wannier model reveals that the lowest energy orbitals in the model are the Bi lone pair states, located around -10 eV below the Fermi level. Nonetheless, due to their bonding interactions with the oxygen p states, they do play a role at the top of the valence band.

\begin{table}
    \centering
    \begin{tabular}{|c|c|c|c|c|} \hline  
         &  \multicolumn{2}{|c|}{\ce{BiOCl}}& \multicolumn{2}{|c|}{\ce{Bi2YO4Cl}}\\ \hline  
         Space Group &  \multicolumn{2}{|c|}{P4/nmm}& \multicolumn{2}{|c|}{P4/mmm}\\ \hline  
 & Wyckoff Position& Point Group& Wyckoff Position&Point Group\\ \hline Bismuth& 2c &4mm ($C_{4v}$)& 2h  &4mm ($C_{4v})$\\ \hline Chlorine& 2c &4mm ($C_{4v}$)& 1b &4/mm ($D_{4h}$)\\ \hline  Oxygen& 2a &$\bar{4}$2m ($D_{2d}$)& 4i &4mm ($C_{4v}$)\\ \hline  Yttrium&  --&--& 1a &4/mm ($D_{4h}$)\\ \hline 
    \end{tabular}
    \caption{Symmetry Data for \ce{BiOCl} and \ce{Bi2YO4Cl} from Materials Project Crystal structures\cite{BiOClMaterialsProject,Bi2YO4ClMaterialsProject,MaterialsProject,SymmetryCharacterTables}.}
    \label{tab:symmetrydata}
\end{table}
\begin{table}
    \centering
    \begin{tabular}{|r|c|c|} \hline 
         &  \ce{BiOCl}& \ce{Bi2YO4Cl}\\ \hline 
         All Hoppings On&  2.6095 eV& 1.4416 eV\\ \hline 
         All Hoppings Off&  3.2374 eV& 3.4807 eV\\ \hline 
         Just Bi Hoppings&  2.3980 eV& 1.6859 eV\\ \hline 
         Just Cl Hoppings&  2.8956  eV& 3.5692 eV\\ \hline 
         Just O Hoppings&  2.7603 eV& 2.3617 eV\\ \hline 
         Just Y Hoppings&  --& 3.4014 eV\\ \hline 
         Just Lone Pair Hoppings&  2.4083 
 eV&  1.8365 eV\\ \hline 
 All Hoppings but Lone Pairs& 3.0608 eV& 1.4603 eV\\ \hline
 All Hoppings but Bi& 2.7532 eV&2.3615 eV\\ \hline 
 All Hoppings but Cl& 3.1510 eV&1.4675 eV\\\hline\hline
 All Hoppings but Y& --&1.5099 eV\\\hline
 All Hoppings but Bi + Y& --&2.3615 eV\\\hline
    \end{tabular}
    \caption{Band Gap Energy Data for \ce{BiOCl} and \ce{Bi2YO4Cl} after diagonalization procedures, with different orbitals connected/disconnected.}
    \label{tab:BandGaps}
\end{table}

We can quantify the effects of the lone pairs in the system by switching the inter-orbital hoppings 'on' or 'off.' By turning all orbital hoppings off, such that only the onsite energies and same-orbital hopping from one unit cell to the other (which are nearly 0) are left. This is represented in the Projected Density of States (PDOS) plots in Figures \ref{fig:PDOS} and \ref{fig:PDOS2} . By turning on only lone pair hoppings, we found that lone pairs mix with either oxygen 2p or chlorine 3p orbitals to create bonding and antibonding orbitals. This mixing pushes the bonding orbital, primarily comprised of the Lone Pair, lower in energy and the antibonding orbital, primarily comprised of the Oxygen or Chlorine orbital; higher in energy to raise the Valence Band Maximum (VBM); all the bands are also broadened. As a result, the band gap energy is lowered. The mixed bonding and antibonding orbitals for each material are shown in Fig \ref{fig:Bonding}. The  band gap energy decreases from 3.2374 eV to 2.4083 eV for \ce{BiOCl} and 3.4807 eV to 1.8365 eV for \ce{Bi2YO4Cl} as we add lone pair hoppings back after turning all orbital hoppings off off. With our methodology, we can obtain a similar picture by turning off lone pair hoppings from when all hoppings are turned on. This has an inverse effect by raising the band gap energy and lowering the VBM.  We analyze the effect of connecting/disconnecting certain orbitals in table \ref{tab:BandGaps}. Both materials experience a sharp effect on the band gap due to the lone pair bonding with the oxygen p states. This directly demonstrates that the lone pairs in these systems are chemically active, contributing to the overall electronic structure of each of these materials.

Photoluminescene in this class of materials is a result of a relaxation from excited electrons in the Bi p states in the conduction band to the Bi lone pair states at the top of the valence band. Despite a large energy gap between the Bi p states and the predominantly Bi lone pair bands at ~10eV below the Fermi level, the Bi lone pair character at the top of the valence band enables this transition (Fig. \ref{fig:BandsDetailed}).
Layered oxyhalides containing transition metal ions, such as \ce{Bi2YO4Cl}, and related material families, have been studied for their photocatalytic properties.  An important aspect of their functionality is that, unlike in \ce{BiOCl}, the top of the valence band in these materials is primarily composed of oxygen (O) p-states rather than chlorine (Cl) states. 
This leads to one of the key differences between \ce{BiOCl} and \ce{Bi2YO4Cl}: their relative photostability under photo-excited chemical reactions. \ce{BiOCl} and the larger family, \ce{BiOX} (X = Cl, Br, I),  slowly degrades over time while \ce{Bi2YO4Cl} and its larger family remains photostable\cite{Photostability}. The standing theory for this difference in photostability is that photo-oxidation in \ce{BiOCl} occurs according to Equation \ref{eqn:Photooxidation} after a positive hole is formed from photo-excitation. This is hypothesized to occur in \ce{BiOX} materials due to the halides' proximity to the top of the VBM, while in \ce{Bi2YO4Cl} or similar materials, the bulk of the halide states are lower in energy in the valence band, making this photo-oxidation less prevalent.

 \begin{equation}
\ce{X^-} + \ce{h^+} \longrightarrow \frac{1}{2}\ce{X2}
\label{eqn:Photooxidation}
\end{equation}
In both \ce{Bi2YO4Cl} and \ce{BiOCl}, the O p-states exhibit more covalent behavior than the Cl states, as their broader bandwidths show. Our findings indicate that Cl states bond less with other orbitals, and remain higher in energy in \ce{BiOCl} compared to \ce{Bi2YO4Cl}. Without hopping, Cl states would dominate the top of the valence band in both materials, a picture which can be understood as resulting from local crystal field splitting, and Madelung potentials (which other authors have discussed). However, in \ce{Bi2YO4Cl}, the combination of more covalent O states and a lower Madelung energy difference between Cl and O states causes the valence band to be dominated by O p-states. 
The results of this analysis are similar to those of other researchers\citeauthor{Bi2MO4ClLP}, which showed that the Bismuth lone pair comprises part of the VBM, allowing us to confirm these findings while simultaneously visualizing lone pair position, energy, and effect on the electronic structure using a standardized procedure\cite{Bi2MO4ClLP}.

Separately, our analysis finds that the role of Y d states is relatively minor, as can also be seen in their effect on the band gap in Table \ref{tab:BandGaps}. The role of the Y atom in this material is then confined to its effect on the crystal structure via ionic radius, and ionic charge state.

Based on these results, we have shown that this method is generalizable not only to lone pairs, but any specific atomic orbital set to understand their role in the overall electronic structure. Similar modeling could easily be extended for the larger family of multimetal bismuth oxyhalides of the type \ce{Bi2MO4Cl}, M ={Y, Gd, La} to examine how changing the transition metal affects the electronic structure or properties between members of this family.

\section{Methodology Details}

\begin{figure}
    \centering
    \includegraphics[width=1\linewidth]{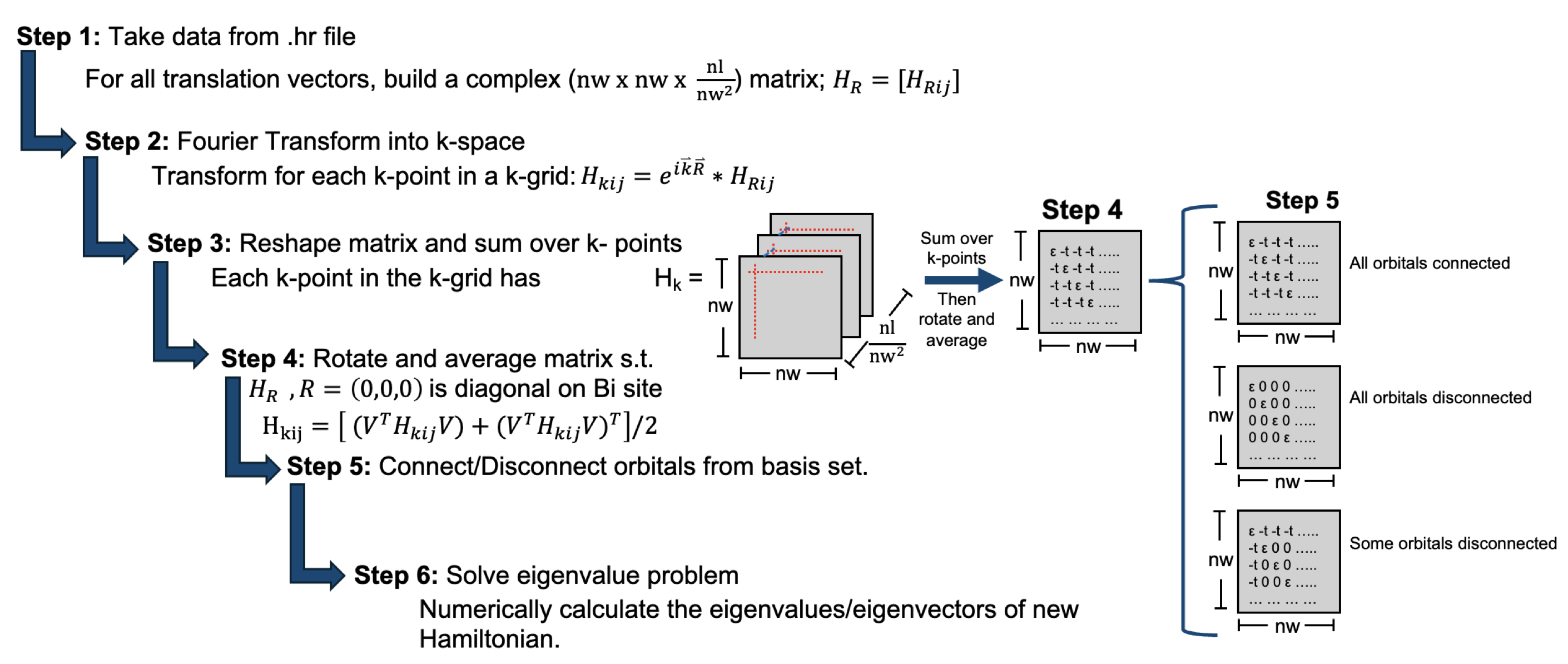}
    \caption{Example of the $\vec{R}=[0,0,0]$ component of the Hamiltonian on one of the Bismuth atoms, before (top) and after (bottom) the rotation used to obtain the Wannier orbitals in this paper for \ce{BiOCl} (left) and \ce{Bi2YO4Cl} (right). The Fermi level for \ce{BiOCl} is 8.09eV while for \ce{Bi2YO4Cl} it's 9.17eV. }
    \label{fig:Ham}
\end{figure}
\begin{figure}
    \centering
    \includegraphics[width=1\linewidth]{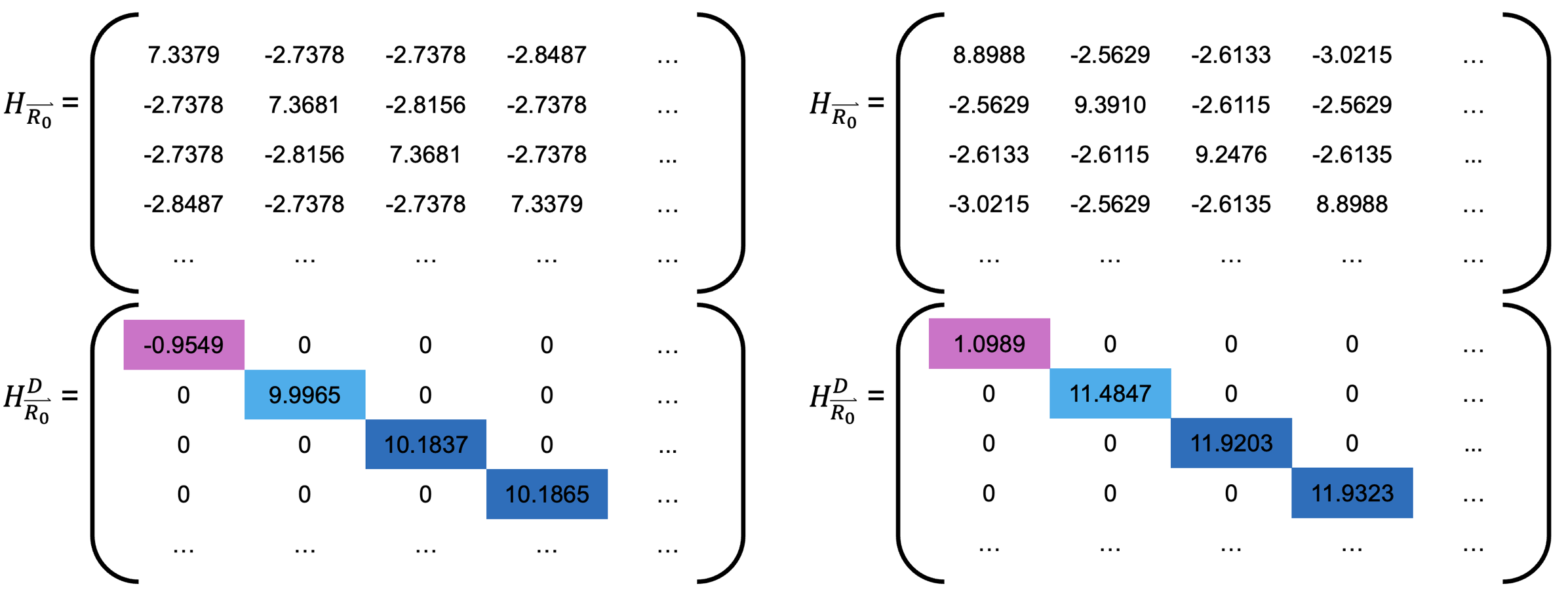}
    \caption{Outline of the steps performed after obtaining the MLWF from Wannier90 to obtain the results presented in this paper.}
    \label{fig:methodology}
\end{figure}

Our procedure begins by performing DFT calculations using the Quantum Espresso\cite{QE1,QE2} code, using the  Perdew-Burke-Ernzerhof (PBE\citep{PBE}) exchange correlation functional, ultrasoft pseudopotentials\citep{Ultrasoft}, and a K-mesh of 9 x 9 x 9 and energy cutoff of 884.4 eV. Following a full crystal structure relaxation, we build a Wannier tight-binding model that fully reproduces the original DFT band structure within an energy range from  -5.0eV to 23.0 eV - where the the Fermi level for \ce{BiOCl} is 8.09eV while for \ce{Bi2YO4Cl} it's 9.17eV, using the Wannier90 code\citep{wannier90}.
This method, outlined in Figure \ref{fig:methodology} and previously discussed, begins with generating Wannier functions using the MLWF methodology and their corresponding Hamiltonian using the Wannier90 software in such a way that the resulting Hamiltonian fully reproduces the bands within the chosen energy range around the Fermi level. We then obtain 3D plots of the Wannier wavefunctions and the Hamiltonian for the material. After obtaining this information, we utilized Matlab scripts to load the Hamiltonian file into matrix form and process it. We provide these tools on Github\cite{Github}. 

The scripts take the Hamiltonian, transform into reciprocal (k) space, and then average over values to transform it into a nw by nw matrix, where nw is the number of orbitals in our Wannier function basis. Using a matrix that diagonalizes, the local $\vec{R}$ = [0, 0, 0] Hamiltonian matrix on certain atoms - in this case the orbitals on the Bi ion, where $\vec{R}$ is a vector that describes the distance in unit cells from one atom to another, we can rotate the Hamiltonian to obtain a more physically meaningful basis, as previously demonstrated on 2D Halide materials \citep{PrevLPABG}. $\vec{R}$ = [0, 0, 0] corresponds to the same unit cell, while for example $\vec{R}$ = [1, 0, 0] corresponds to hopping from one unit cell to the nearest neighbor one along the $\vec{a}$ axis. Using the vectors from the rotation and the initial Wannier functions, we generate 3D plots of the new Wannier functions with VESTA, allowing us to visualize the corresponding linear superposition of Wannier orbitals in a diagonal on-site basis.

We can switch off-diagonal hopping values "on" or "off" by setting those values to zero while keeping the on-site energy values unchanged. It is important to note that we never entirely disable same-orbital hoppings between unit cells, resulting in a small but finite bandwidth, even when all other orbital hoppings are turned off. Additionally, disconnecting all orbitals and gradually turning hoppings on, versus starting with all hoppings active and then selectively disconnecting them, are two distinct processes that are not reversible. For example, if we start with all hoppings active and disconnect the lone pairs, the columns and rows corresponding to the lone pair hoppings are set to zero. Conversely, starting with no hoppings and activating all orbitals except the lone pairs results in all hopping-related columns and rows being active, producing a complete Hamiltonian.

The eigenvalues and eigenvectors derived from the new Hamiltonian—regardless of how it is obtained—can then be used to generate new PDOS (Projected Density of States) and projected orbital band plots for further analysis. At any stage, orbital interactions can be toggled on or off, leading to the analysis discussed in the previous section.

\section{Discussion}
Our Wannier function-based approach builds on standard methods by introducing two additional steps. The first step involves rotating the final set of Wannier orbitals to align them with the local symmetry of the system. The second step selectively toggles on and off the hopping between specific sets of orbitals, allowing us to quantify complex bonding interactions, particularly in materials with a large variety of orbitals and non-zero hopping terms.

This approach enables us to interpret complex bonding in materials using an orthonormal set of states that fully reproduces the DFT results within a certain energy range. It starts with local Madelung and crystal field effects, as reflected in the on-site orbital energies, and extends to analyzing the contribution of each individual bonding interaction between different orbitals and atoms. We have applied this method to \ce{BiOCl} and \ce{Bi2YO4Cl}, which feature lone pairs, multiple anions, and transition metal d-states. This analysis provides insight into the specific role of each orbital in shaping the final electronic structure, demonstrating the wide applicability of the method.

Our methodology can be applied to materials of arbitrary complexity, with different types of relevant local electronic states, using a combination of standard Density Functional Theory (DFT) tools and post-processing scripts.

\begin{acknowledgement}
The authors thank Sara Skrabalak and Varsha Kumari for discussions. EGW was partially supported by the Raymond Siedle Undergraduate Research Fund, the Drs. Sidney and Becca Fleischer Research Scholarship, and the John H. Billman Summer Scholarship. This research was supported by Indiana University startup funds, and in part by the Lilly Endowment, Inc., through its support for the Indiana University Pervasive Technology Institute.
\end{acknowledgement}

\newpage
\bibliography{main}

\end{document}